# Experimental observation of parabolic wakes in thin plates

Janez Rus[1*], Aleksi Bossart[1], Benjamin Apffel[1], Matthieu Malléjac[1] and Romain Fleury[1*]

[1]*Laboratory of Wave Engineering, Institute of Electrical and Micro Engineering, Ecole Polytechnique Fédérale de Lausanne (EPFL), Station 11, 1015 Lausanne, Switzerland*
[*]*Correspondence to romain.fleury@epfl.ch or janez.rus@tum.de*

Wakes are medium perturbations created by a moving object, such as wave patterns behind boats, or wingtip vortices following an aircraft. Here, we report about an experimental study of an uncharted form of parabolic wakes occurring in media with the group velocity twice larger than the phase velocity, as opposed to the conventional case of Kelvin wakes. They are formed by moving a laser spot on a thin plate, which excites a unique wake pattern made of confocal parabolas, due to the quadratic dispersion of the zero-order flexural Lamb mode. If the spatial dimensions are rescaled by the perturbation velocity and material constant, we obtain a single universal wake with constant parabolic focal lengths. We demonstrate an evanescent regime above the critical frequency where the wave components oscillate exclusively in the direction parallel to the perturbation path, with an opening angle of 90°. We define a dimensionless number analogous to Froude and Mach numbers, which determines whether the complete parabolic wake pattern will be excited by the moving source or not.

The broadest definition of wakes refers to a pattern, typically made of waves, excited by the movements of an object or a perturbation in a medium. The most prominent example is the Kelvin wake pattern generated by a moving object on the surface of water [1-11]. A second common use of the term wake is associated with the circulating turbulent flow behind a moving object in a fluid (Kármán vortex) [12,13], as seen with phenomena like wingtip vortices behind an aircraft [14-17]. Additional associations encompass wakes formed behind charged objects in supersonic plasma flows [18-20] or laser-driven plasma wakefields [21-24], which show promise as an alternative method for electron acceleration. This definition also includes Mach cones [19,25] or Cherenkov radiation [26-29], which are less often called wakes due to the absence of dispersion wave interference.

The wake pattern on the surface of deep water depends on whether the dispersion is capillary-dominated [30-32], with $\omega \propto k^{3/2}$ being the relation between angular frequency $\omega$ and wavenumber $k$, or if it is gravity-dominated [8,33], with $\omega \propto k^{1/2}$. The latter power law yields the commonly-observed Kelvin wake pattern, typical of the wakes originating from ducks or boats as they move. The Kelvin wake patterns exhibit the remarkable property that their shape, in the case of sufficiently slow movements, is independent of the perturbation velocity, when properly rescaled coordinates are used. The half-cone opening angle is constant at the value $\arcsin(1/3) \cong 19.47°$. This is a common situation of the wakes on water where the group velocity is twice lower than the phase velocity.

In this work, we experimentally explore the behavior of wake patterns excited in a medium in which the power law governing the dispersion relation is exactly inverse to the one of Kelvin wakes, namely $\omega \propto k^2$. These wakes are obtained behind a fast-moving perturbation on a thin plate. The zero-order asymmetric Lamb mode of the plate exhibits a dispersion relation well approximated by $\omega = k^2/\alpha$, where $\alpha$ is a constant dependent on the thin plate thickness and material properties. The group velocity $c_\mathrm{g} = \mathrm{d}\omega/\mathrm{d}k = 2k/\alpha$ is therefore twice higher than the phase velocity $c_\mathrm{ph} = \omega/k = k/\alpha$, and proportional to the square root of frequency, setting a unique stage for the study of the new form of wakes.

The perturbation in our case is a focused laser beam illuminating the surface of a polymer plate (Fig. 1a). The induced heating leads to a localized reduction in the Young's modulus of the plate. In our experiment, galvanometric scanning mirrors were used to displace the heated position at constant velocities $v_\mathrm{p}$ in a linear trajectory. The angle between the plate surface and the laser beam was set to 20° at the midpoint of the scanning area. Consequently, a small change in the galvanometric mirror's angle provided a large alteration in the perturbation's position. This, in turn, allowed us to achieve perturbation velocities $v_\text{p}$ fast enough to observe a broad range of parabolic wakes. Moreover, this setup enabled us to investigate the impact of the laser spot size on the formation of the parabolic wakes. For more details regarding the experimental procedure and the setup, please refer to Supplemental Material [34].

In Fig. 1 b-g, we present the parabolic wakes captured by a laser vibrometer for three distinct $v_\text{p}$ values and two different time instances. The out-of-plane velocity of the plate was measured for each scanning position through separate wake excitations. The excitation laser was in focus at the position $x_\parallel = 0$, corresponding to time $t = 0$ (Fig. 1 b, d, and f). Additional wakes for 12 different $v_\text{p}$ values and a scenario involving an elevated temperature of the polymer plate are presented in Supplemental Material and Video Presentation [34].

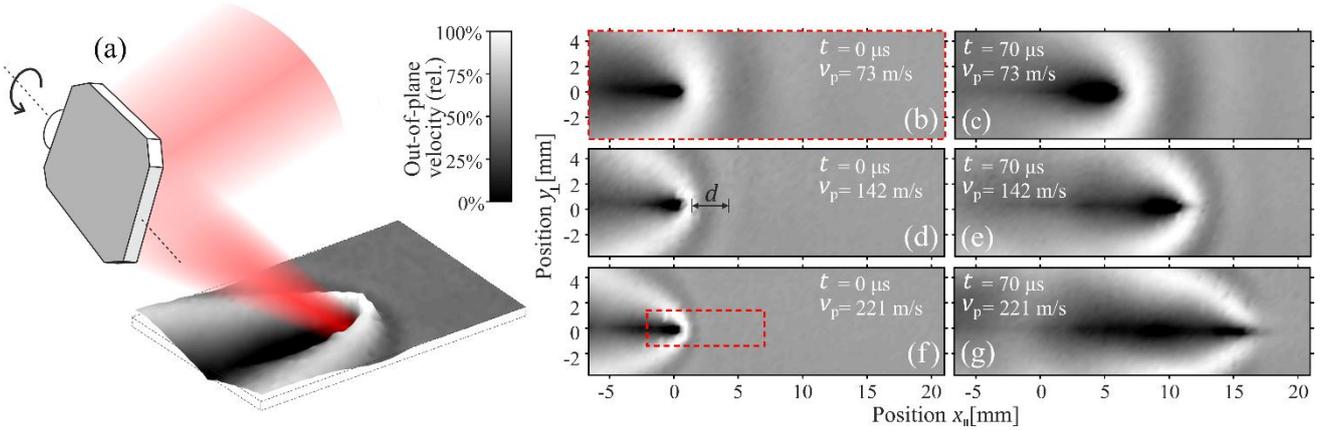

Fig. 1. Schematic of the setup, where a moving laser beam excites elastic waves on a thin plate (a). Measured parabolic wake patterns at t = 0 μs (b, d, f) and t = 70 μs (c, e, g) at three distinct perturbation velocities $v_\text{p}$ (b-c, d-e, and f-g). The red dashed-line square in f shows a portion of the wake whose shape is identical to the one observed over the full range in b, when both axes are rescaled by a factor three, corresponding to the ratio of $v_\text{p}$ values in these two cases.

The wave pattern shape of Fig. 1b is identical to that confined within the red dashed-line square in Fig. 1f, which has both axes reduced by a factor 3 – chosen to correspond to the ratio between the values of $v_\text{p} = 221$ m/s in Fig. 1f and $v_\text{p} = 73$ m/s in Fig. 1b. In contrast, Kelvin wakes follow an inverse rescaling law, where higher $v_\text{p}$ provide larger dominant wavelengths, while rescaling of spatial coordinates in both directions does not change the cone opening angle. In our case, the measured parabolas all have a curvature that linearly increases with $v_\text{p}$. The focal distance, which is inversely proportional to the curvature of the parabolic ridges in Fig. 1b-g, is therefore inversely proportional on $v_\text{p}$.

As wakes result from the interference of many frequency components, we propose to study the different spectral components of the wakes separately. As examples, we show in Fig. 2d-f (respectively Fig. 2g-i) the measured scans from Fig. 1b,c at $v_\text{p} = 73$ m/s (respectively from Fig. 1f,g at $v_\text{p} = 221$ m/s), filtered to three selected narrowband frequencies, which were then processed to obtain data points in Fig. 2a-c, marked by yellow squares (respectively green circles).

In our experiment (Fig. 2a), we observe that the wave number of the wakes in the direction parallel to the motion is fixed by $v_\text{p}$, and follows the law

$$k_\parallel(\omega) = \frac{\omega}{v_\text{p}}. \qquad (1)$$

In particular, it is neither influenced by the dispersion law of the medium nor the wave velocities in the medium. Instead, its dependency on $\omega$ is entirely determined by $v_\text{p}$. This condition fixes for each frequency the phase of the travelling wave at the moving location of the source. Equivalently, it matches the parallel projection of the phase velocity with the source velocity. Using the dispersion law $\omega = k^2/\alpha$ where $k = \sqrt{k_\parallel^2 + k_\perp^2}$, we can express the wave number of the wakes in the direction perpendicular to $v_\text{p}$:

$$k_\perp(\omega) = \sqrt{\alpha\omega - k_\parallel^2} = \sqrt{\alpha\omega - \frac{\omega^2}{v_\text{p}^2}} = \alpha v_\text{p} \sqrt{\frac{\omega}{\alpha v_\text{p}^2} - \frac{\omega^2}{\alpha^2 v_\text{p}^4}}. \qquad (2)$$

If we introduce rescaled variables $\hat{k}_\perp = k_\perp/\alpha v_\text{p}$ and $\hat{k}_\parallel = k_\parallel/\alpha v_\text{p}$, as well as $\hat{\omega} = \omega/\alpha v_\text{p}^2$, we can simplify Eq. (1) and Eq. (2) to

$$\begin{aligned}\hat{k}_\parallel(\hat{\omega}) &= \hat{\omega} \\ \hat{k}_\perp(\hat{\omega}) &= \sqrt{\hat{\omega} - \hat{\omega}^2}\,.\end{aligned} \qquad (3)$$

The above formula for rescaled wavenumbers $\hat{k}_\parallel(\hat{\omega})$ and $\hat{k}_\perp(\hat{\omega})$ is represented in Fig. 2a and Fig. 2b with thick red lines and compared to the experimental results obtained for ten different source velocities ranging from 73 m/s to 292 m/s identifiable by the shades of gray. The opening angle between the propagation direction of the wave component and the direction perpendicular to the source motion at a specific frequency, $\varphi(\hat{\omega}) = \arctan\left(\hat{k}_\parallel(\hat{\omega})/\hat{k}_\perp(\hat{\omega})\right)$, is also shown in Fig. 2c. As a result of the rescaling, the darker shades of gray, corresponding to higher $v_\text{p}$, stop at lower values of $\hat{\omega}$ in Fig. 2a-c, even though the range of non-rescaled $\omega$ is constant for all $v_\text{p}$. Fig. 2a-c in their original (non-rescaled) coordinates are provided in Supplemental Material [34].

Eq. (3) predicts that $\hat{k}_\perp(\hat{\omega})$ is real for $\hat{\omega} < 1$, when $\omega$ is smaller than the critical frequency $\omega_\text{cr} = \alpha v_\text{p}^2$. The angle $\varphi(\hat{\omega})$ increases until it reaches 90° at $\omega_\text{cr}$. This is the highest frequency at which the wake components are purely propagating. The wake component at $\omega_\text{cr}$ propagates in the direction parallel to the perturbation path with the velocity equal to $v_\text{p}$. It can therefore be directly expressed as $\omega_\text{cr} = v_\text{p}/d$, where $d$ is the distance between the vertices of two parabolas defined by the maximum ridges of the wake pattern (Fig. 1d). Interestingly, an opposite behavior is true for Kelvin wakes. The $k_\perp(\omega)$ of Kelvin waves is real above a critical frequency, which can be estimated from the wavelength of the wave components propagating directly behind the moving perturbation.

For parabolic wakes, $\hat{k}_\perp(\hat{\omega})$ is imaginary at $\hat{\omega} > 1$. In this evanescent regime, wave components are localized to the vicinity of the perturbation path, oscillating only along the direction parallel to $v_\text{p}$, as governed by Eq. (1). This effect is evident in Fig. 2f. The exponential decay length in the direction perpendicular to $v_\text{p}$ of the evanescent waves decreases with increasing frequencies. Due to the limited frequency range of the measuring system and the increase of intrinsic losses at higher frequencies, the evanescent regime ($\hat{\omega} > 1$) is more easily observed in the measurements performed at lower $v_\text{p}$.

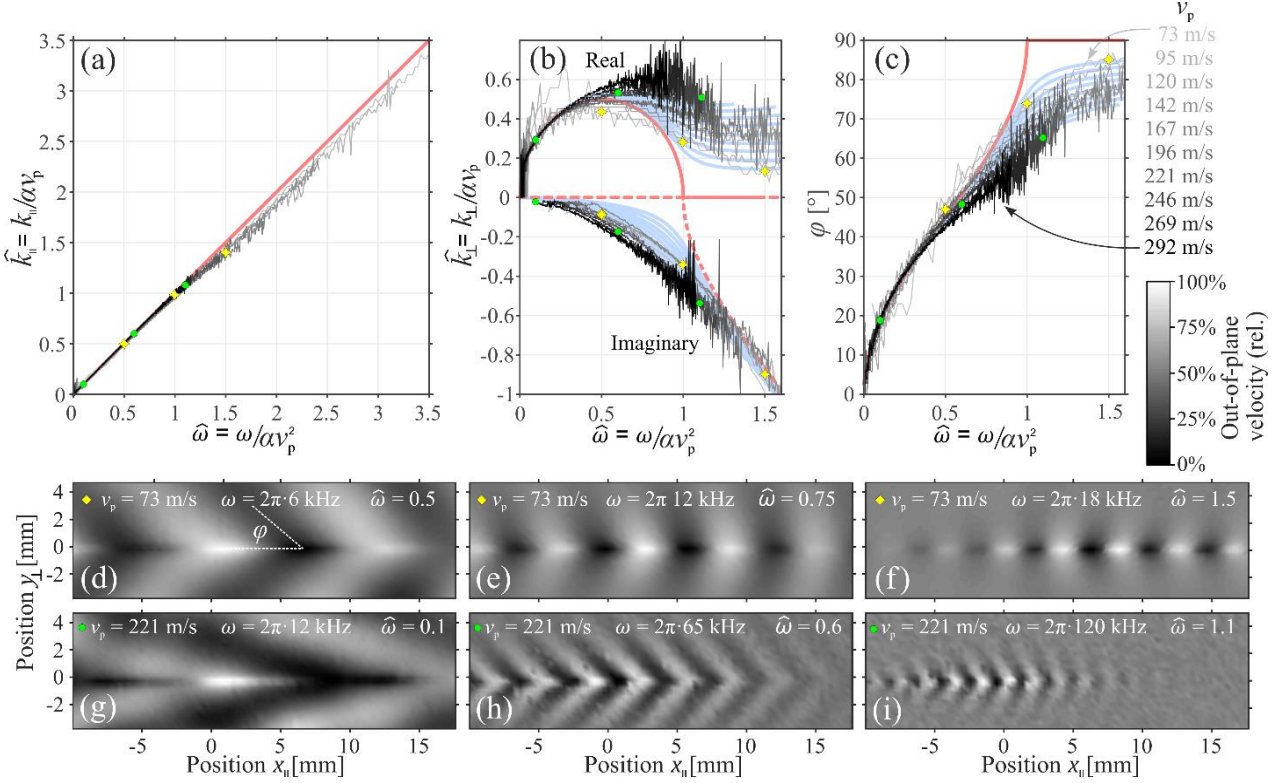

Fig. 2. The measured wave numbers $\hat{k}_\parallel$ (a) and $\hat{k}_\perp$ (b) for ten different $v_p$ values (grey lines), collapse on the red curves predicted by Eq. (3) except near the critical frequency (at $\hat{\omega} = 1$). Around the critical frequency, variations in the perturbation size along the position $x_\parallel$ (due to the laser beam moving in and out of focus) become important. These variations are accounted for in a refined model (blue lines). The angle $\varphi$ between the direction of $v_p$ and the wavefronts becomes larger as $\hat{\omega}$ increases, reaching almost 90° at the critical frequency $\hat{\omega} = 1$ (c). Beyond this frequency, Eq. (3) predicts that $\hat{k}_\perp$ becomes imaginary, which defines the evanescent regime (f). When the parabolic wake patterns from Fig. 1 are filtered to narrow frequencies (d-i), represented by yellow squares and green circles, the openings of the wavefronts adhere to the Mach law in respect to the wave velocity at the specific frequency.

While $\hat{k}_\parallel$ falls on the same curve for all $v_p$ values, there is a greater discrepancy between the measured $\hat{k}_\perp$ and Eq. (3), especially around $\omega_{cr}$. This arises due to variations in laser beam diameter as it moves along in the direction of $v_p$. As the beam approaches the focal point, higher frequency components are stimulated with higher amplitudes (and conversely as it moves away from the focal point). These alterations of the excitation amplitude spectrum along $x_\parallel$ leads to a non-zero imaginary part of $\hat{k}_\parallel$. As a result, the imaginary part of $\hat{k}_\perp$ becomes non-zero even below the critical frequency, and the real part of $\hat{k}_\perp$ is underestimated by Eq. (2) when nearing $\omega_{cr}$ (Fig. 2b). This effect becomes more pronounced for larger $v_p$ values. Consequently, for the field map of Fig. 2i, the angle $\varphi$ is 65° (green point in Fig. 2c), instead of 90° as predicted by the theory. The effect of the laser focusing and defocusing along the position $x_\parallel$ can be accounted for in a refined model (Supplemental Material [34]), yielding the thick blue lines, which fit the measurements better than the reduced model of Eq. (3).

Having understood how each frequency behaves individually, we can now express the wake pattern as the real part of the following sum over all excited waves that verify Eq. (1) and Eq. (2)

$$Z(x,y,t) = \int_0^{\omega_{Max}} A(\omega) e^{-ik_x x} e^{-ik_y y} e^{-i\omega t} d\omega = \int_0^{\omega_{Max}} A(\omega) e^{-i\frac{\omega}{v_p}x} e^{-i\sqrt{\alpha\omega - \frac{\omega^2}{v_p^2}}y} e^{-i\omega t} d\omega. \quad (4)$$

$\omega_{Max}$ represents the highest frequency being excited, while $A(\omega)$ is the excitation amplitude at a given $\omega$. The amplitude spectrum of the measured parabolic wake is provided in Supplemental Material [34] for 12 different perturbation velocities $v_p$.

Using the same rescaled variables as those employed for Eq. (3), the out-of-plane displacement of the interference wake pattern can be formulated in rescaled position coordinates: $\hat{x}_\parallel = x_\parallel \alpha v_p$, $\hat{y}_\perp = y_\perp \alpha v_p$, and rescaled time $\hat{t} = t\alpha v_p^2$ as:

$$Z(\hat{x}_\parallel, \hat{y}_\perp, \hat{t}) = \int_0^{\hat{\omega}_{Max}} A(\hat{\omega}) e^{-i\hat{\omega}\hat{x}} e^{-i\sqrt{\hat{\omega}-\hat{\omega}^2}\,\hat{y}} e^{-i\hat{\omega}\hat{t}} d\hat{\omega} \qquad (5)$$

Analogously to the case of Kelvin wakes, the integral solution can be expressed using the stationary phase approximation (please refer to Supplemental Material [34]). In order to explain the parabolic wake pattern of the velocity (out of the plane $x_\parallel$-$y_\perp$), which is what we measure, we need to differentiate Eq. (5) with respect to $\hat{t}$. Fig. 3a shows the stationary phase approximation of the time derivative of Eq. (5) at $\hat{t} = 0$ and absolute $\hat{y}_\perp$ values. The two-dimensional interference pattern has ridges and zero values that conform to the shape of parabolas. All their focal points are located at the current position of the perturbation $(\hat{x}_\parallel, \hat{y}_\perp) = (0,0)$. The focal lengths of the $n$ confocal parabolas, (i.e. the distance between their vertices and the current position of the perturbation) lying on the maximum ridges of the wake pattern, are given by $\widehat{f_n} = \phi_p + \pi/4 + 2\pi n$ in $(\hat{x}_\parallel, \hat{y}_\perp)$ coordinates. $\phi_p$ is a global constant phase shift depending on the details of the excitation physics and $n$ is an integer labeling the considered parabola. For the out-of-plane velocity measured in our experiment, a phase $\phi_p$ equals to $\pi/2$ is expected and observed, since the maximum plate displacement of the wake field occurs at the point illuminated by the heating laser. These statements were cross-validated by numerical integral solutions and analytical stationary phase approximation as detailed in Supplemental Material [34].

We now propose a geometrical construction of the parabolic pattern (Fig. 3b). For this, one can draw parallel lines representing the wavefronts of specific narrowband frequency components (similar to the examples from the measurement in Fig. 2d-i). These lines have inversely signed slopes (or angle $\varphi$ in Fig. 2d-i) for positive and negative $\hat{y}_\perp$, representing Mach cones with the symmetry line on the $x_\parallel$ axis. The relation between these slopes and the distance between two neighboring lines is defined by the dispersion relation ($c_{ph}$ at specific frequency component), while the origin of the lines – intersection with the ordinate axis – is defined by $\phi_p$. The lines are tangent to confocal parabolas with focal lengths $\widehat{f'}_n = \phi_p + 2\pi n$, as shown in Supplemental Material [34]. This approach also allows us to recover the wake patterns associated with other dispersion relations, for example to compare with Kelvin wakes (see Video Presentation). When the line bundles are replaced by narrowband frequency components (with the lines following the maxima of the two-dimensional sloped harmonic functions), a pattern similar to Fig. 3a emerges. The parabolas situated on the maximum ridge of this wake pattern are more open and have larger foci, shifted by $\pi/4$. This is because all the tangent lines fall outside the parabolas with focal lengths $\widehat{f'}_n = \phi_p + 2\pi n$.

The focal lengths of the confocal parabolas were measured experimentally by determining the position $x_\parallel$ of the maximum signal amplitude (and the first minimum signal amplitude in the positive direction from the maximum) for all the positions $y_\perp$, for all time instances when the parabolic wake pattern was within the scanned region, and for ten different $v_p$. The obtained mean values of the position $x_\parallel$ along with their standard deviation bands in rescaled coordinates $(\hat{x}_\parallel, \hat{y}_\perp)$ are presented in Fig. 3c. The curves obtained at all ten different speeds (coded in gray shades) and at the increased α (blue) align with two confocal parabolas having focal lengths $3\pi/4$ (maximum) and $3\pi/4 + \pi$ (minimum). This provides validation for our mathematical models and the universality of the parabolic wake pattern.

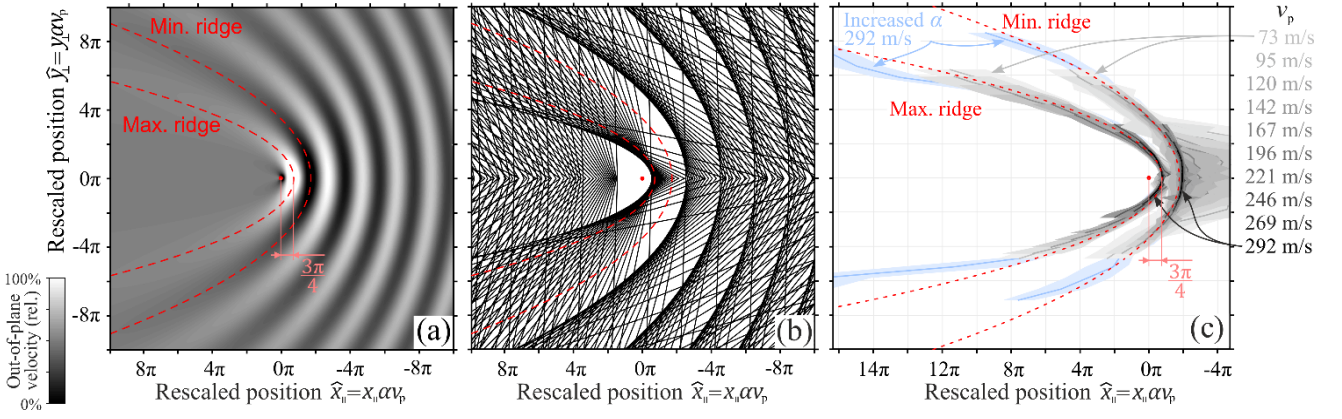

Fig. 3. Time derivative of Eq. (5) obtained by the stationary phase approximation (a). The maximum ridges of the pattern follow shapes of confocal parabolas, having focal lengths $\widehat{f}_n = \phi_p + \pi/4 + 2\pi n$, where $\phi_p = \pi/2$ due to the velocity measurement. The parabolic shape of the interference pattern can be elucidated through the concept of parallel lines (wavefronts) (b). Their slope (or angle $\varphi$) is computed from their periodicity (wavelength) using the dispersion relation. Their origin corresponds to the phase of the wake excitation. If the dimensions $x_\parallel$ and $y_\perp$ are rescaled by the factor $\alpha v_p$, the maximum and minimum ridges observed in the measurement (full wakes at three chosen speeds shown in Fig. 1) align with the two confocal parabolas (red dashed lines) with focal lengths of $3\pi/4$ and $3\pi/4 + \pi$, respectively, across all ten different $v_p$ (c, gray shades) as well as for the scenario when $\alpha$ is increased (c, blue).

To achieve the complete parabolic wake pattern, it is essential that the moving perturbation excites frequencies reaching at least until $\omega_{cr}$. This condition is not fulfilled when the length of the perturbation spot $L$ is excessively large or $\alpha v_p$ is overly high. In such cases, the parabolic wake pattern starts to open from the front, where the waves depart from the perturbation at $\varphi = 90°$. Due to the lack of the frequency components with larger $\varphi$ (having the wider cone opening), the parabolic wake pattern tends to approximate the shape of a Mach cone. This situation is similar to the transition between Kelvin and Mach regimes, wherein the cone openings are narrower than the Kelvin angle if low-frequency components are not excited, for instance in the case of excessively small objects moving on a water surface with too high velocity [6,7].

In analogy with the Mach number for non-dispersive media and the Froude number for Kelvin wakes, a dimensionless number that governs these physics can be defined for media with quadratic dispersion, as $R = v_p \alpha L$. This number measures the ratio between the perturbation velocity and the critical velocity $v_{cr} = 1/\alpha L$. The wake pattern has its complete shape around the vertices of parabolas when $R \leq 1$ (Fig. 1b-f). All frequency components up to $\omega_{cr}$ are excited if $L$ is sufficiently small at specific $v_p$. This is not the case in Fig. 1g. Since the laser is out of focus, the value of L is too high. At lower values of $v_p$, the criterion $R \leq 1$ is achieved for smaller $L$. In this case, the parabolas of the wake pattern will be closed at their vortices, however, they will have a smaller slope (longer focal lengths). In other words, only limited area around the central part of the rescaled universal wake pattern will be visible.

To summarize, we provided an analytical explanation for experimentally observed parabolic wakes propagating in a medium with quadratic dispersion. We have shown that the observed pattern is universal once the coordinates are rescaled by the velocity and material factors. The equations define two regimes (propagating and evanescent) separated by $\omega_{cr}$. In opposition to the Kelvin wake, the wave components below $\omega_{cr}$ produce the parabolic wake, while evanescent wave behavior is observable under the condition that the addressed frequency $\omega \geq \omega_{cr}$. This condition is reminiscent to the critical angle behavior in the phenomenon of total reflection. The evanescent waves propagate solely along the trajectory of the moving perturbation, instead of along a spatial interface for the phenomenon of total reflection.

Our findings can be extended to other phenomena governed by quadratic (or even other power laws) dispersion, such as flexural phonons on graphene membranes [35-37] and specific regimes of polaritons in semiconductor microcavities [38-40].

We also demonstrated that Lamb waves can be generated not only by a laser pulse as a fast temporal change in illuminating power, but also by swift spatial movements of a continuous laser beam. This phenomenon holds potential for applications in contact-free damage detection and imaging of mechanical properties that influence the shape of the parabolic wake.

In our experiment, the wave propagation properties were altered by inducing changes in the heat distribution at the spot of the moving perturbation. Interesting wave phenomena are anticipated to emerge as a consequence of interaction between the moving perturbation and the wake pattern that was excited at a prior temporal instance (similar to a study on water waves [41]). This situation occurs when the trajectory of the moving perturbation deviates from a straight path, when $v_p$ is not constant, or when the intensity of the perturbation varies over time.

# Supplemental Material:
# Experimental observation of parabolic wakes in thin plates

Janez Rus[1*], Aleksi Bossart[1], Benjamin Apffel[1], Matthieu Malléjac[1] and Romain Fleury[1*]

[1]*Laboratory of Wave Engineering, Institute of Electrical and Micro Engineering, Ecole Polytechnique Fédérale de Lausanne (EPFL), Station 11, 1015 Lausanne, Switzerland*
*Correspondence to romain.fleury@epfl.ch or janez.rus@tum.de*

## 1. Experimental methods

The parabolic wakes were generated using a continuous laser (FL-1064-CW, manufactured by Changchun New Industries Optoelectronics Technology) with a power of 6 W and a wavelength of 1064 nm (Fig. S1b). A 2-axis galvanometer scan head (XG210, manufactured by Mecco) was employed to move the laser spot continuously along the surface of the thin plate at a velocity $v_\mathrm{p}$ (Fig. S1c).

The thin plate with a thickness of 0.1 mm was constructed from a shape memory polymer supplied by SMP Technologies Inc, Tokyo. The polymer exhibited a glass transition temperature within the range of 25°C to 90°C (Fig. S1a). Its Young's modulus decreases considerably, by a factor of at least 20, when its temperature is increased several tens of degrees Celsius above room temperature [1]. Consequently, the wave propagation properties are rapidly altered upon exposure to the moving continuous laser. This perturbation gave rise to the formation of parabolic wake patterns. The thin plate was cooled by a steady flow of room-temperature air flow directed to the specimen surface using a nozzle (Fig. S1e).

The ultrasonic responses of the thin plate in the form of the parabolic wake patterns were independently detected for each scanning positions (pixels) in Fig. 1 and Fig. S2. This was feasible due to the high repeatability of the wake excitation. To achieve this, a PSV-F-500-HV laser scanning vibrometer (Polytec) was installed on the side of the thin plate opposite to the continuous laser (Fig. S1d). For each scanning position, the wake pattern was stimulated 10 times with a repetition rate of 20 Hz to achieve sufficient signal averaging. The sampling frequency was set at 6.25 MHz. The surface of the thin foil illuminated by the wake exciting laser was coated black to enhance absorption, whereas the opposing side, illuminated by the laser of the vibrometer, was covered with a retro-reflective foil.

In order to test the hypothesis that the shape of the parabolic wake in rescaled coordinates remains constant even if $\alpha$ changes as well, the temperature of the thin plate was raised. Uniform heating across the entire scanned area was achieved by positioning a heater at a distance of 1 cm from the thin plate. The local temperature was elevated by 5°C, resulting in a 68% increase in $\alpha$, which was measured independently as detailed in Section 6.

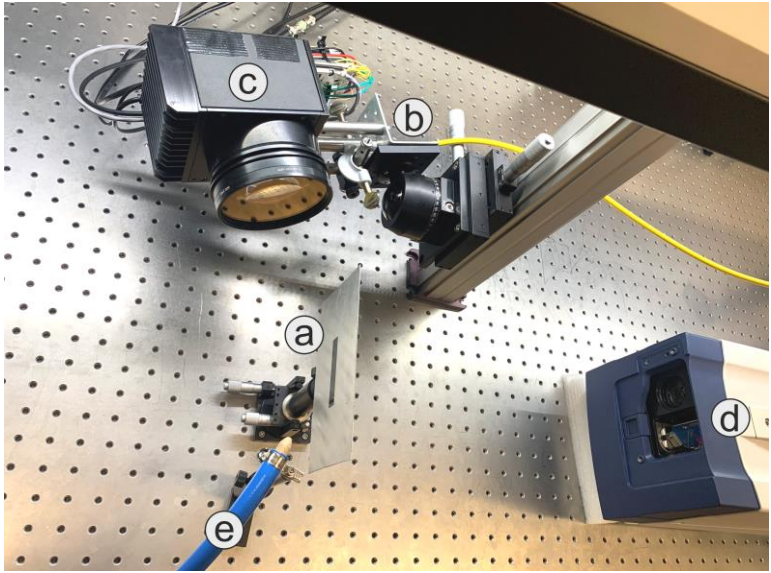

Fig. S1. Parabolic wakes propagated within a medium, which was a thin black polymer plate mounted in a metal frame (a). The moving perturbation was induced by a fiber-guided continuous laser (b) directed through a galvanometric scanning head (c). The wakes were measured utilizing a laser scanning vibrometer (d). The sample was cooled by an air nozzle (e).

## 2. Parabolic wake pattern measurement at twelve speeds and increased temperature

The scans of the wake patterns were performed over $89 \times 29$ positions ($x_\parallel$, $y_\perp$) with a spatial resolution of 0.31 mm. The zero value of the position $y_\perp$ was set to the line of the wake excitation (the symmetry line of the parabolic wake). Similarly, the zero value of the position $x_\parallel$ was defined where the continuous laser came into focus on the thin plate's surface.

In Fig. S2, we present two timeframes of the measured wake patterns for 12 distinct $v_\text{p}$ values and the scenario when the temperature (and $\alpha$) was increased (Fig. S2y). The first timeframe (left column) corresponds to $t = 0$ μs, marking the moment when the perturbation crosses $x_\parallel = 0$ mm. This instant aligns with the moving laser being in focus (perturbation size is the smallest). The second time frame (right column) corresponds to the time instant when the perturbation crosses $x_\parallel = 12$ mm. Here, the laser is out of focus, resulting in a limited range of the excited frequency components. Consequently, the shapes of the parabolic wake patterns undergo changes at higher $v_\text{p}$ values.

Signals shown in Fig. S2e-y were used to generate Fig. 2 and Fig. 3c. The analysis in Supplemental Material encompassed all 12 different $v_\text{p}$ values.

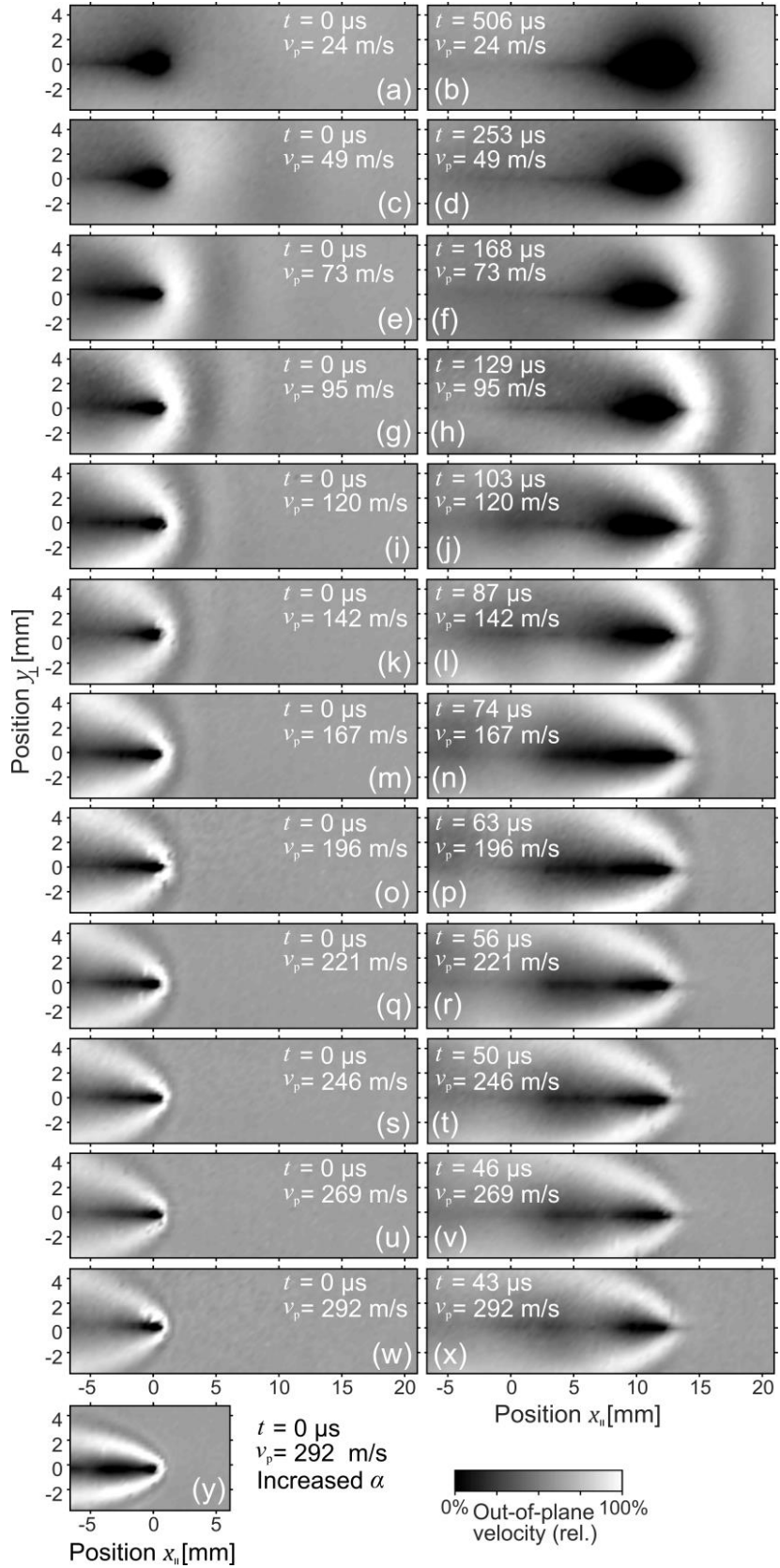

Fig. S2. When the laser beam is in focus (perturbation at $x_\parallel = 0$, left-hand column), high frequency components are excited. In this scenario, two maximum ridge parabolas become evident, with their vertex distance governed by $\omega_{cr}$: this distance is larger for lower $v_p$ and vice versa. However, when the laser beam is out of focus (perturbation at $x_\parallel = 12$, right-hand column), we can only observe the second maximum ridge parabola up until $v_p = 167$ m/s (n). For higher dimensionless numbers $R \leq 1$ (being a result of higher $v_p$ values), the parabolic wakes open at the vertices and approach the Mach regime. Please refer to the Video Presentation for the complete animation.

## 3. Wavenumber-frequency diagrams without coordinate rescaling

The upper half of the wake patterns (encompassing the 15 highest positions $y_\perp$) were utilized to obtain Fig. 2 and Fig. S3. Two positions along the line of the wake excitation at $y_\perp = 0$ mm were excluded from the analysis. In the initial step, the signals from all scanning positions and all $v_p$ were transformed to the frequency domain using a fast Fourier transform. Subsequently, an exponential function $Ae^{Bx} + C$ was fitted to the absolute values of the obtained frequency components for all scanning positions and all $v_p$. Here, the constant $B$ represented the imaginary part of $k_\perp$.

The real parts of $k_\parallel$ and $k_\perp$ were determined by the following procedure. A two-dimensional fast Fourier transform was performed on the images across the dimensions $x_\parallel$ and $y_\perp$ for each of the frequency components (examples in Fig. 2d-i) and for all $v_p$. This operation yielded images with two peaks situated at positions $(k_\parallel, k_\perp)$ and $(-k_\parallel, -k_\perp)$, due to the central point symmetry.

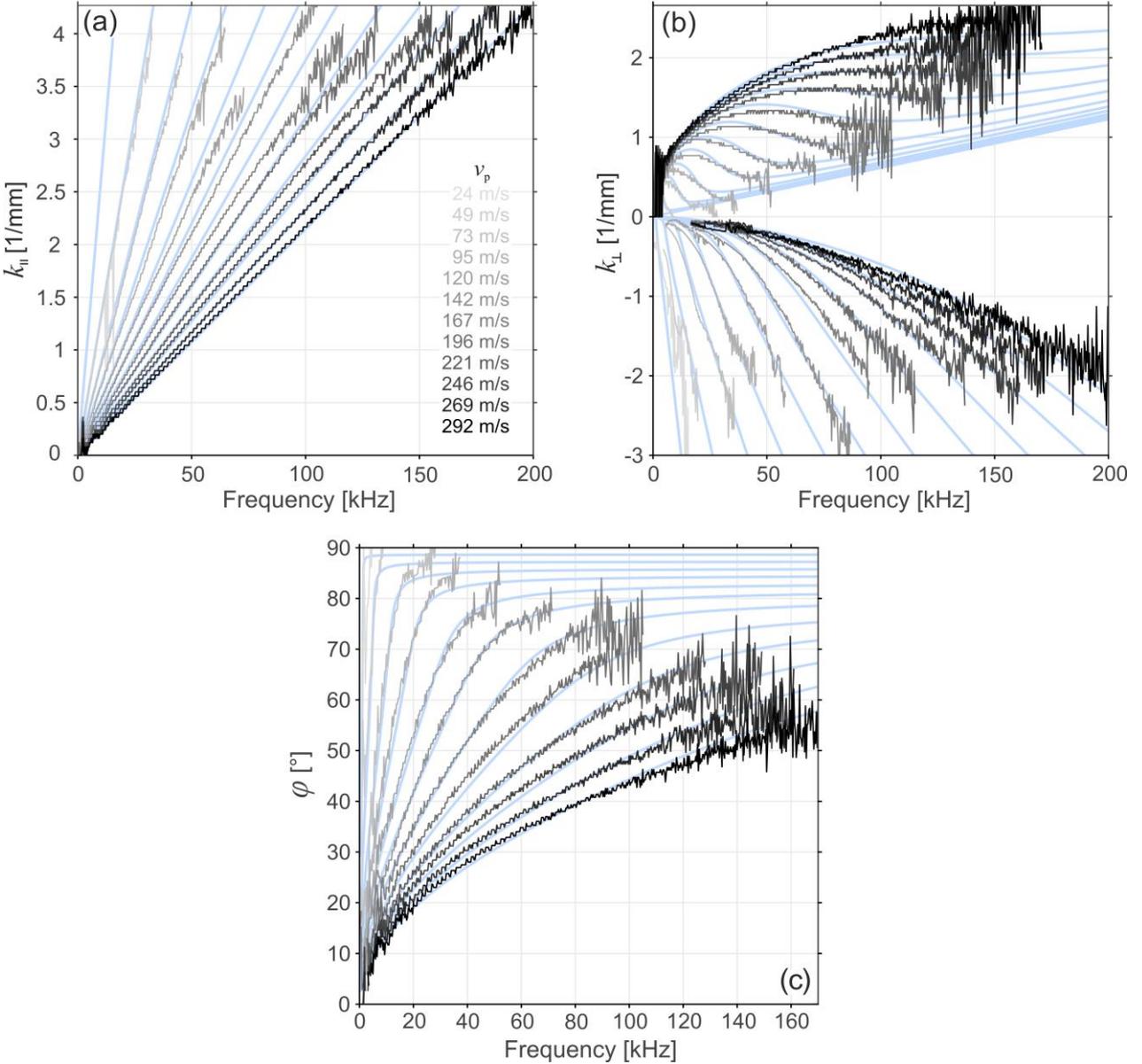

Fig. S3. The data of Fig. 2a-c (with additional two measurement at lower $v_p$) are presented with the difference that the coordinates are not rescaled: wave numbers $k_\parallel$ and $k_\perp$, as well as the opening angle $\varphi$ in relation to the frequency ($\omega/2\pi$). As a result, the measured curves (gray shades) no longer overlap, but instead conform to the values anticipated theoretically by Eq. (1) and Eq. (2). The imaginary part of $k_\parallel$ is included, which arises due to the variation in the laser spot size along the direction $x_\parallel$ (Section 10). The red curve displayed in Fig. 2 is omitted from this representation, as it differs for each $v_p$ when coordinates are not rescaled.

The imaginary part of $k_\parallel$ exerts an influence resulting in elevated values of the real part of $k_\perp$ particularly for higher frequencies (as elaborated further in Section 10). The imaginary part of $k_\perp$ is non-zero even prior to reaching the critical frequency $\omega_{cr}$. As a result, the trajectory of $k_\perp$ curve is smooth in the vicinity of $\omega_{cr}$. The angle $\varphi$ is reduced due to the influence of the imaginary part of $k_\parallel$. This phenomenon stems from the effect that as the laser approaches its focal point, progressively higher wave amplitudes are stimulated.

## 4. Frequency spectrum excited by the moving perturbation

The measurement of the signal amplitudes close to the trajectory line of the perturbation (0.6 mm above the symmetry line of the parabolic pattern), reveals the presence of two distinct spectral shape regimes (Fig. S4). In the first regime, encompassing the lowest six $v_p$, a noticeable change of the amplitude slope (in respect to the frequency) can be observed. This change aligns with the corresponding $\omega_{cr}$ (gray-scale vertical lines in Fig. S4a, black vertical line in Fig. S4b). This phenomenon arises from the fact that the frequency components above $\omega_{cr}$ do not propagate in the far field, as clarified by Fig. 2 (evanescent behavior). In the second regime, encompassing the highest six $v_p$, there is no pronounced change of the amplitude slope (in respect to the frequency). This is due to the intrinsic losses that increase with frequency ($\omega_{cr}$ features a quadratic dependence on $v_p$), and the influence of the imaginary part of $k_\perp$ (Section 10), which becomes more prominent at higher $v_p$.

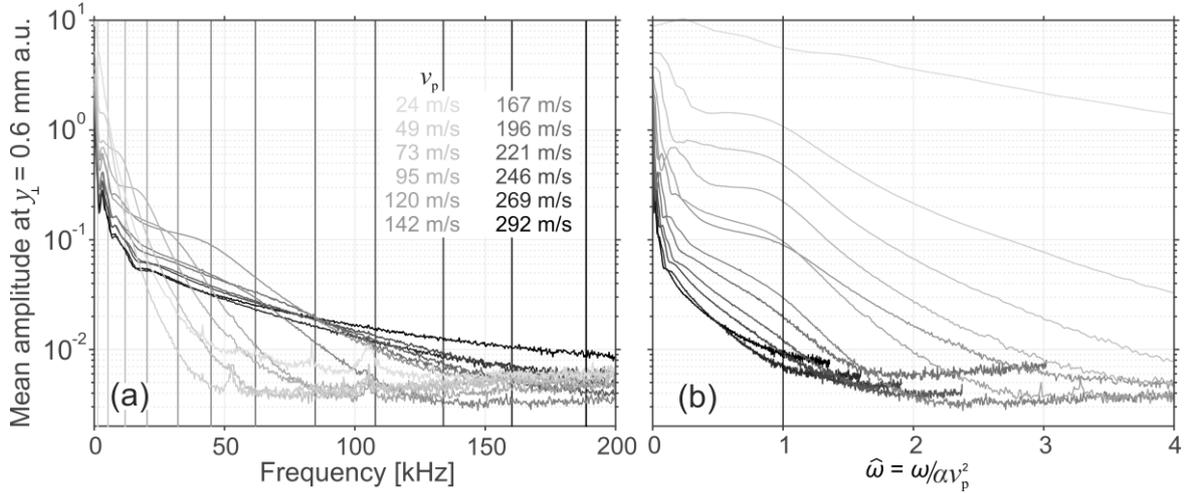

Fig. S4. The mean amplitude of the excited wake pattern, encompassing all $x_\parallel$ positions at $y_\perp = 0.6$ mm is presented for 12 distinct $v_p$ values in the original frequency coordinate (a), as well as in the rescaled angular frequency coordinate (b).

## 5. Measured focal lengths of confocal parabolas and difference between the maximum and minimum ridge parabolas

As explained in the main part of the article (Fig. 3), the wake pattern in the medium with quadratic dispersion has the shape of confocal parabolas. Rescaling both spatial dimensions (e.g. parallel and perpendicular to the velocity direction of the moving perturbation), by the factor $\alpha v_p$, consistently yields focal lengths $\widehat{f}_n = \phi_p + \pi/4 + 2\pi n$, where $n$ denotes the integer relating to the specific parabola situated along the maximum ridge of the wake pattern.

For each time instance in which the parabolic wake was within the scanning range, we pinpointed the locations $x_\parallel$ with maximum and minimum signal amplitudes for all $y_\perp$ values and all $v_p$ variations (including the scenario with increased $\alpha$). This search procedure was applied on unfiltered signals. We defined the condition that minimum signal amplitudes were positioned at higher $x_\parallel$ values compared to maximum signal amplitudes. This was necessary in order to exclude the areas characterized by deeply negative amplitude values after the perturbation.

Following this procedure, maximum and minimum ridge parabolas shown in Fig. 3c were derived. Subsequently, parabolic curves ($x_\parallel$ of minimum or maximum amplitudes in dependency on $y_\perp$) were fitted for all time instances and all $v_p$ values. The focal lengths were determined from the parabolas' slope (from the constant of the quadratic term) and then rescaled by $\alpha v_p$. Mean and standard deviation of these measured focal lengths are displayed in Fig. S5a with a minimum of 319 values used for each $v_p$ value and for both foci – with more values at lower $v_p$, due to the longer presence of the wake pattern within the scanning range. For the experiment involving increased $\alpha$, 119 values were used. The measured focal lengths closely align with the values delivered by our analytical model: $\hat{f}_n = 3\pi/4$ for the maximum ridge parabola and $\hat{f}_n = 7\pi/4$ for the minimum ridge parabola. Please note that in the equation $\hat{f}_n = \phi_p + \pi/4 + 2\pi n$, $n = 0$ for the first maximum ridge parabola and $n = 1/2$ for the first minimum ridge parabola. In our experiment, $\phi_p$ equated to $\pi/2$ since we measured velocity of the out-of-plane displacement of the thin plate ($\phi_p$ would be 0 in the case of a displacement measurement).

Fig. S5b graphically presents the distances between the vertices of the fitted maximum and minimum ridge parabolas of the measurement. If these distances are scaled by the factor $\alpha v_p$, they closely align with the value of $\pi$ as predicted by our analytical model.

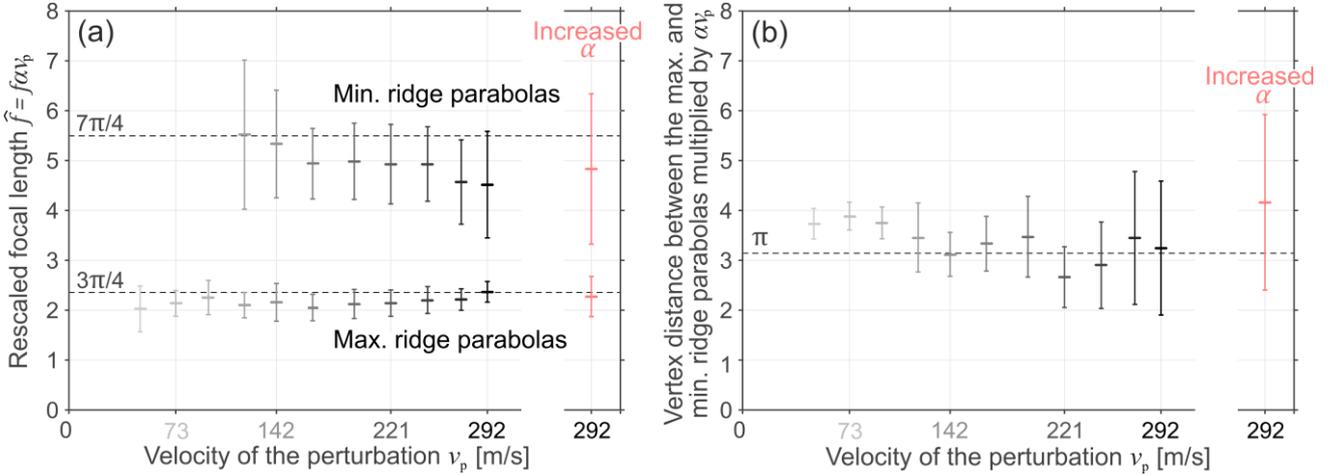

Fig. S5. The focal lengths of the first maximum and minimum ridge parabolas, derived directly from the experimental data depicted in Fig. S2, exhibit alignment with the values of $3\pi/4$ (maximum ridge parabola) and $7\pi/4$ (minimum ridge parabola) when scaled by the factor $\alpha v_p$. The measured distances between the vertices of both parabolas closely approximate the value of $\pi$ under the same rescaling law.

## 6. Measurement of the material constant $\alpha$

The material constant $\alpha$, which depends on the material and thickness of the thin plate, was determined through a separate measurement. In this procedure, ultrasound was excited using a laser pulse with the wavelength of 532 nm, energy of 10 mJ, duration of 5 ns (full width at half-maximum), and a repetition rate of 20 Hz utilizing a Surelite SL I-20 laser (Manufacturer: Continuum). The laser pulse beam diameter was 5 mm (95% energy). A linear scan was executed across 111 positions $x$ (aligned parallel to $x_\parallel$) with a spatial resolution of 0.31 mm, extending away from the ultrasound source. This scan covered the same region of the thin plate as utilized for the measurement of parabolic wake patterns.

Measured raw time signals in Fig. S6a reveal a distinct presence of a slower zero-order asymmetric Lamb wave, characterized by its typical dispersive shape. A faster zero-order symmetric Lamb wave mode, discernible as a narrow straight bright line, departing from $t = 0$ with near-horizontal slope, was excluded from the $\alpha$ estimation.

The time-position diagram presented in Fig. S6a was converted to $\omega$-$k$ diagram by a two-dimensional fast Fourier transform. In subsequent steps, we pinpointed the value of $k$ corresponding to the maximal amplitude of the zero-order symmetric Lamb wave component for each $\omega$. The $\alpha$ value was quantified by fitting a square root function over the measured $k(\omega)$ relationship. For the scenario involving room temperature (black curve in Fig. S6a), we obtained the value $\alpha = 13.8 \text{ s/m}^2$, and for the scenario of increased thin plate temperature $\alpha = 23.17 \text{ s/m}^2$ (red curve in Fig. S6a, Fig. S2y).

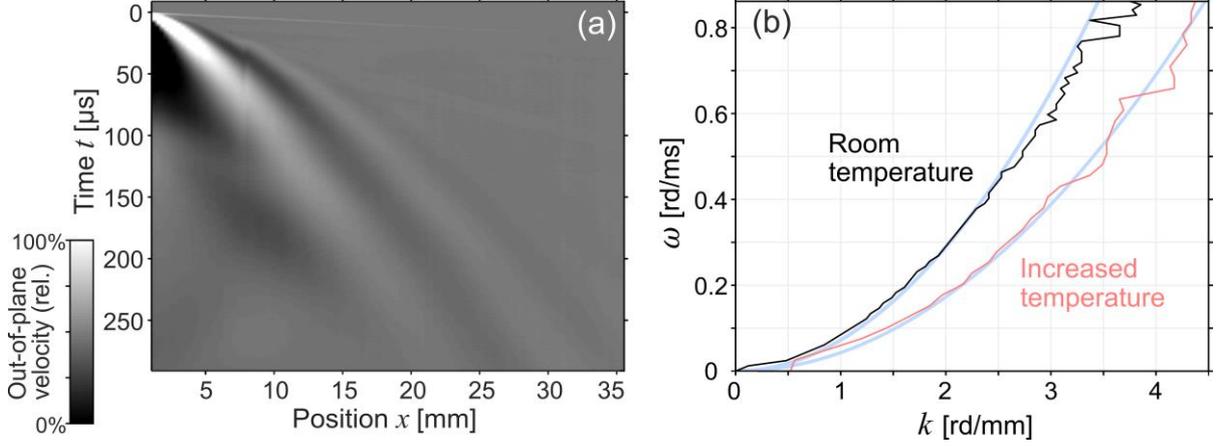

Fig. S6. The unprocessed signals acquired through a linear scan along the position $x$ (aligned parallel to $x_\parallel$) away from the pulse excitation point at $x = 0$ (a). $\omega$-$k$ diagrams of the zero-order asymmetric Lamb wave (maximum amplitude values) at room temperature (b, black curve) and at the plate temperature increased for 5°C (b, red curve). For both cases, α values were determined by fitting a square root function to the measured $k(\omega)$ data.

## 7. Solving the wake pattern integral by stationary phase approximation

In the co-moving frame, employing the vectorial representation for the perturbation velocity $\vec{v}_\text{p} = v_\text{p}\vec{x}_\parallel$, the quadratic dispersion relation can be expressed as follows:

$$\omega(k) = k^2/\alpha - \vec{k} \cdot \vec{v}_\text{p}. \qquad \text{Eq. S1}$$

The stationary wave pattern in polar coordinates in the co-moving frame generated by the moving perturbation can be represented as a summation of numerous plane waves with zero frequency:

$$Z_\text{CF}(\vec{r}) = \int A(\vec{k}) e^{i(\omega t - \vec{k}\cdot\vec{r})} \delta\left(\omega(\vec{k})\right) d\vec{k}. \qquad \text{Eq. S2}$$

We introduce a vectorial expression for $\vec{k} = k(\cos\theta\vec{x}_\parallel + \sin\theta\vec{y}_\perp)$, where $\theta$ is the complementary angle of the cone opening angle $\varphi$ at specific frequency (Fig. 2d-i).

By writing $\vec{k} = k(\cos\theta\vec{x}_\parallel + \sin\theta\vec{y}_\perp)$ one can rewrite $\omega = 0$ as

$$k = v_\text{p}\alpha\cos\theta. \qquad \text{Eq. S3}$$

As $k > 0$, this forces $-\pi/2 < \theta < \pi/2$. We also write $\vec{r} = r(\cos\phi\,\vec{x}_\parallel + \sin\phi\vec{y}_\perp)$ so that

$$\vec{k}\cdot\vec{r} = kr\cos(\theta - \phi). \qquad \text{Eq. S4}$$

Additionally, we will make the assumption that the emission amplitude is homogeneous $A(k,\theta) = A_0$. The integral can now be written as

$$Z_\text{CF}(\vec{r}) = \int_{k=0}^{\infty}\int_{\theta=0}^{2\pi} A(k,\theta)\,e^{-i\vec{k}\cdot\vec{r}}\delta(k - v_\text{p}\alpha\cos\theta)\,k\,\text{d}k\,\text{d}\theta$$
$$= \int_{-2\pi}^{2\pi} A_0\,e^{-iv_\text{p}\alpha r\cos(\theta)\cos(\theta-\phi)} v_\text{p}\alpha\cos(\theta)\text{d}\theta \qquad \text{Eq. S5}$$

$$= A_0 v_p \alpha \int_{-\pi/2}^{\pi/2} \cos(\theta)\, e^{-i\hat{r}\cos(\theta)\cos(\theta-\phi)}\, d\theta$$

$\hat{r} = r v_p \alpha$ is rescaled by the same factor, as discussed in the main part of our work. This integral is invariant under the transformation $\phi \to -\phi$. Consequently, we can further assume that $0 < \theta < \pi$.

As the integral of Eq. S5 is in the form of $\int f(\theta) e^{-i\hat{r}g(\theta)}\, d\theta$, it can be approximated using stationary phase approximation when $r \to \infty$. For this purpose, one seeks the stationary points of $g$ such that $g' = -\sin(2\theta_c - \phi) = 0$.

Considering the constrains on $\theta$ and $\phi$, the two stationary points are

$$\theta_1 = \frac{\phi}{2} \quad \text{and} \quad \theta_2 = \phi/2 - \pi/2. \qquad \text{Eq. S6}$$

One can consequently solve the integral describing the parabolic wake pattern in co-moving frame as

$$Z_{CF}(\vec{r}) = A_0 v_p \alpha \sqrt{\frac{\pi}{\hat{r}}} \left[ \cos(\phi/2) e^{i(-\hat{r}\cos^2(\phi/2)+\pi/4)} + \sin(\phi/2) e^{i(\hat{r}\sin^2(\phi/2)-\pi/4)} \right]. \qquad \text{Eq. S7}$$

Remember that this expression holds true solely within the range of $0 < \theta < \pi$. When extending the range to $\pi < \theta < 2\pi$, it becomes necessary to apply absolute values to both the cosine and sine terms. This step ensures that the integral maintains its invariance and yields the correct expression for all $\phi$.

Within our study, the measured quantity is the out-of-plane velocity, as opposed to the out-of-plane deformation in the co-moving frame, which is the case for the derivation in this section until Eq. S7. To align with our experimental conditions, we formulate the parabolic wake pattern in the stationary frame. This is achieved through a coordinate transformation:

$$Z(\vec{r}, t) = Z_{CF}(\vec{r} - \vec{v}_p t). \qquad \text{Eq. S8}$$

The out-of-plane velocity in polar coordinates can then be written (assuming that the deformation of the thin plate for the whole parabolic wake pattern remains small) as

$$\dot{Z}(\vec{r}, t) = \partial_t Z(\vec{r}, t) = -\vec{v}_p \cdot \vec{\nabla} Z_{CF}. \qquad \text{Eq. S9}$$

Utilizing the integral form of $Z_{CF}(\vec{r})$ as described by Eq. S5 and taking into account the perturbation velocity $\vec{v}_p = v_p \vec{x}_\parallel$, we can formulate $\dot{Z}(\vec{r}, t)$ at $t = 0$:

$$\dot{Z}(\vec{r}) = -i A_0 v_p^3 \alpha^2 \int_{-\pi/2}^{\pi/2} \cos^2(\theta) \cos(\theta - \phi)\, e^{-i\hat{r}\cos(\theta)\cos(\theta-\phi)}\, d\theta \qquad \text{Eq. S10}$$

Through a computation using the stationary phase approximation, a process akin to the one employed for deriving Eq. S7, we arrive at the expression for the parabolic wake pattern taking in account the experimental conditions

$$\dot{Z}(\vec{r}) =$$
$$-i A_0 v_p^3 \alpha^2 \sqrt{\frac{\pi}{\hat{r}}} \left[ |\cos(\phi/2)|^3 e^{i(-\hat{r}\cos^2(\phi/2)+\pi/4)} + |\sin(\phi/2)|^3 e^{i(\hat{r}\sin^2(\phi/2)-\pi/4)} \right]. \qquad \text{Eq. S11}$$

$\dot{Z}(\vec{r})$ provides a parabolic pattern symmetric with regard to two Cartesian axes as it includes sum of causal and anti-causal cases. In order to obtain the pattern observed in the experiment (symmetric only with regard to perturbation's trajectory), we consider the causal case alone – the first term of the equation (before the plus symbol). This delivered us the pattern presented in Fig. 3a.

## 8. Geometric derivation of the parabolic wake pattern

In this section, we derive the shape of the caustics depicted in Fig. 3b of the main text. There, waves are approximated as trains of lines defined by the condition $\vec{k} \cdot \vec{r} = 2\pi n$, where $n$ is an integer. Assuming a general power-law dispersion $\omega(k) = k^p$ and using the direct equivalent of Eq. (3) of the main text, this condition yields

$$x = \frac{2\pi n}{\widehat{\omega}} - y\sqrt{\frac{\widehat{\omega}^{\frac{2}{p}}}{\widehat{\omega}^2} - 1}. \qquad \text{Eq. S12}$$

To find the caustic, we must find a point along this line that remains fixed under small variations in $\widehat{\omega}$, as shown in Fig. S7a. In particular, the $x$ coordinate of this fixed point does not change under a small variation in $\widehat{\omega}$, yielding the condition $\partial_{\widehat{\omega}} x = 0$. Together with Eq. S12, this allows us to solve for the coordinates of the caustic point associated to the angular frequency $\widehat{\omega}$, namely

$$x = \frac{2\pi n}{p-1}(p\widehat{\omega}^{1-\frac{2}{p}} - \widehat{\omega}^{-1})$$

$$y = p\frac{2\pi n}{p-1}\sqrt{\widehat{\omega}^{-\frac{2}{p}} - \widehat{\omega}^{2-\frac{4}{p}}} \qquad \text{Eq. S13}$$

This parametric expression contains the Kelvin-wake case ($p = 1/2$), the capillary-wave case ($p = 3/2$), and the case treated in the main text ($p = 2$). As shown in Fig. S7b, it also allows us to recover the asymptotic pattern corresponding to $p$ tending towards infinity. In our case of interest, $p = 2$, we can go further and remove $\widehat{\omega}$ from Eq. S13. As expected, we obtain parabolic caustics satisfying

$$y = 4\pi n\sqrt{1 - \frac{x}{2\pi n}}. \qquad \text{Eq. S14}$$

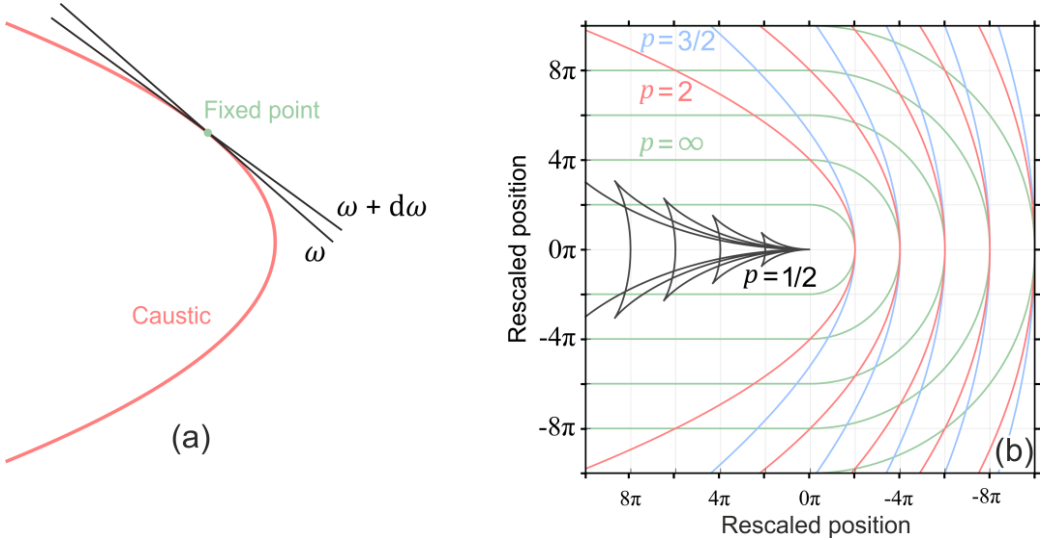

Fig. S7. (a) Two wave fronts (black lines) that are close in frequency interfere constructively at a fixed point, depicted in green. We use this fixed-point condition to derive caustic wave patterns for arbitrary power laws, with the $p = 1/2, 3/2, 2, \infty$ cases depicted in black, blue, red, and green respectively (b).

## 9. Proof that $\phi_p = \pi/2$ for the velocity measurement

As a starting point, we take Eq. (5) of the main part of the article, wherein $Z(\hat{x}_\parallel, \hat{y}_\perp, \hat{t})$ represents the out-of-plane displacement induced by the parabolic wake pattern. The numerical solution of its integral is presented in Fig. S8a, corresponding to $\hat{t} = 0$.

In the first case, we assume a phase shift $\phi_p = \pi/2$ for all frequency components, consequently leading to the inclusion of the term $e^{-i\frac{\pi}{2}}$ in Eq. (5):

$$Z_{\phi_p=\pi/2}(\hat{x}_\parallel, \hat{y}_\perp, \hat{t}) = \int_0^{\hat{\omega}_{Max}} A(\hat{\omega}) e^{-i\hat{\omega}\hat{x}} e^{-i\sqrt{\hat{\omega}-\hat{\omega}^2}\,\hat{y}} e^{-i\hat{\omega}\hat{t}} e^{-i\frac{\pi}{2}} d\hat{\omega}. \qquad \text{Eq. S15}$$

The numerical solution of Eq. S15 at $\hat{t} = 0$ is presented in Fig. S8b. One can observe that in comparison to Fig. S8a, the parabolas traveling along the maximum (and minimum) ridge of the pattern exhibit longer focal lengths when $\phi_p = \pi/2$. The dashed red lines signify the confocal parabolas with foci $\hat{f}_n = \pi/4 + 2\pi n$, while the solid red lines signify the parabolas with foci $\hat{f}_n = 3\pi/4 + 2\pi n$, where $n$ is an integer.

The numerical solution of the time derivative of Eq. (5)

$$\dot{Z}(\hat{x}_\parallel, \hat{y}_\perp, \hat{t}) = \int_0^{\hat{\omega}_{Max}} -i\hat{\omega} A(\hat{\omega}) e^{-i\hat{\omega}\hat{x}} e^{-i\sqrt{\hat{\omega}-\hat{\omega}^2}\,\hat{y}} e^{-i\hat{\omega}\hat{t}} d\hat{\omega} \qquad \text{Eq. S16}$$

at $\hat{t} = 0$ is shown in Fig. S8c. The maximum ridge parabolas also exhibit focal lengths of $\hat{f}_n = 3\pi/4 + 2\pi n$ as well, since its time derivative introduces a phase shift of $\phi_p = \pi/2$ to all frequency components. This conclusion aligns with a comparison between the parabolic wake patterns described by Eq. S8 and Eq. S11, both of which were derived using the stationary phase approximation. Please note that while the shapes of $Z_{\phi_p=\pi/2}(\hat{x}_\parallel, \hat{y}_\perp, \hat{t})$ and $\dot{Z}(\hat{x}_\parallel, \hat{y}_\perp, \hat{t})$ are identical, their absolute amplitude values differ.

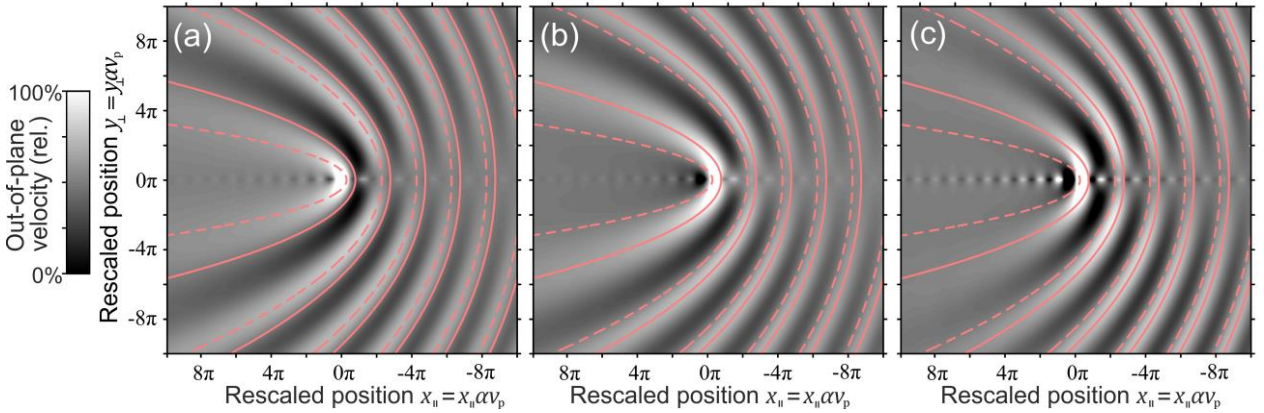

Fig. S8. Introducing a phase shift of $\phi_p = \pi/2$ to all the frequency components (b) yields an equivalent effect on the parabolic wake pattern (a) as performing a time derivation (c). This transition results in a switch of the maximum ridge parabolas from the dashed to the solid red lines – transforming from the family of confocal parabolas with foci $\hat{f}_n = \pi/4 + 2\pi n$ to the family characterized by foci $\hat{f}_n = 3\pi/4 + 2\pi n$, where $n$ is an integer.

## 10. Measurement of the imaginary part of $k_\parallel$

The imaginary part of $k_\parallel$ was a consequence of the laser spot size not being uniform along $x_\parallel$ on the thin plate surface. This non-uniformity was a consequence of the experimental necessity to focus the laser beam to attain sufficient light intensity. Simultaneously, in order to increase the highest achievable $v_p$, the laser beam needed to be inclined to an angle of 20° (in the middle of the scanned region) with respect to the thin plate surface.

The effect of approaching (negative $x_\parallel$) and distancing (positive $x_\parallel$) from the focus along the perturbation's path notably affected $k_\perp$ values near $\omega_{cr}$, as evident by the difference between the blue and red lines in Fig. 2. However, this effect on the parabolic wake pattern (across the entire frequency range) remained relatively limited and could be neglected when assessing the focal distances of the parabolas (Fig. 3c and Fig. S5).

Our objective was to maintain the analytical model as simple as possible and minimize parameters connected specifically to our experimental setup. We thus employed only one constant that characterized the rate of laser focusing along $x_\parallel$. The alteration in the beam size was approximated by an exponential increase and decrease (with the symmetry line at $x_\parallel = 0$), which reasonably approximated the experimental Gaussian variation of laser spot size along $x_\parallel$. This approach resulted in an imaginary component of $k_\parallel$. As it was linked to the laser beam's characteristics, it was unaffected by $v_p$ and linearly dependent on frequency: $\text{imag}(k_\parallel) = D\omega$, with $D = 7.66 \times 10^{-4}$ s/m representing a parameter tied to the laser beam's divergence properties, which remained constant throughout the experiment.

The amplitude spectrum induced by the moving continuous laser exhibited dependency on $x_\parallel$ (averaged values across all $x_\parallel$ are shown in Fig. S4). Higher frequency components were solely excited when the laser (perturbation location) was in focus while traversing the thin plate surface. The laser spot size variation provided an advantage, enabling the observation of the pattern shape at an extended perturbation size where higher frequency components were not excited (right-hand column in Fig. S2 and Fig. 1g).

The imaginary part of $k_\parallel$ can be measured by monitoring the amplitude change of a specific frequency component along $x_\parallel$ (it would be constant if the laser spot size was uniform along $x_\parallel$). In Fig. S9, we present the measurement of exponential decay constants (referring to amplitudes) against frequency for the six highest $v_p$ values, taken from the middle of the upper half of the measured wake patterns ($y_\perp = 2.8$ mm). Other measurements at lower $v_p$ values are omitted due to the low signal-to-noise ratio in the frequency range above 50 kHz.

We can observe that for all $v_p$ values, the measured exponential decay constants (gray lines in Fig. S9) conform closely to the curve defined by the equation: $\text{imag}(k_\parallel) = D\omega$ (blue lines in Fig. S9).

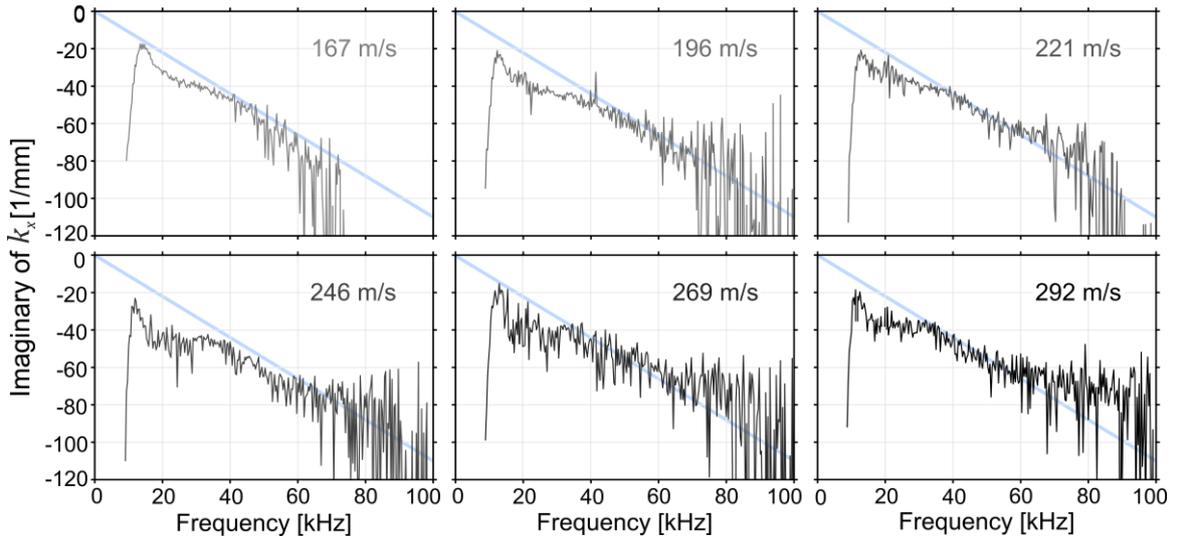

Fig. S9. The gray curves represent the measured exponential decay constants of the amplitude in the $x_\parallel$ direction at $y_\perp = 2.8$ mm. Given that the laser beam was traveling away from its focal spot, the beam diameter expanded at higher $x_\parallel$, leading to the diminished excitation of high-frequency components. This effect is effectively captured by the imaginary part of $k_\parallel$, which remains unaffected by $v_p$ and exhibits linear dependence on frequency.

## 11. Modeled wake pattern including the imaginary part of $k_\parallel$

In Fig. S10, we present the numerical solution of the integral of Eq. -, while considering the non-zero imaginary component of $k_\parallel$. This accounts for the variation in perturbation size along $x_\parallel$ in our experiment. The expression for $k_\parallel$ is given by:

$$k_\parallel(\omega) = \frac{\omega}{v_\text{p}} + iD\omega \qquad \text{Eq. S17}$$

where $D = 7.66 \times 10^{-4}$ s/m has been measured and is positive for $x_\parallel < 0$. As the perturbation travels in the positive $x_\parallel$ direction and the laser beam approaches the focal spot, the perturbation size is contracted and increasingly higher frequency components are excited. The situation is reversed for $x_\parallel > 0$ where $D = -7.66 \times 10^{-4}$ s/m. The numerical example presented in Fig. S10 corresponds to the measured parabolic wake pattern shown in Fig. 1e and Fig. S2q with $v_\text{p} = 221$ m/s and $\alpha = 13.8$ s/m$^2$.

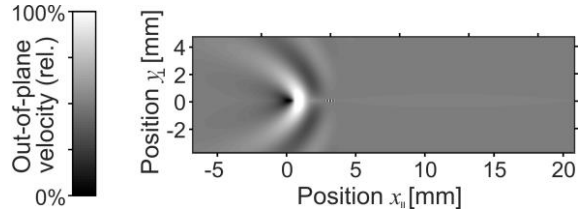

Fig. S10. We consider the variation in perturbation size both in time and along $x_\parallel$, which is a consequence of the perturbation movement towards and away from the position $x_\parallel = 0$, where the laser beam is in focus. At this time instant, the highest frequency components are excited.

## 12. References of Supplemental Material